\documentclass[showpacs,prl,twocolumn,floats,floatfix]{revtex4}

\usepackage{graphicx,epsfig}
\usepackage{times}
\usepackage{bm}
\usepackage{amsmath,amssymb}

\DeclareMathOperator{\im}{Im}

\begin{document}

\title{Quantum-to-classical crossover of
quasi-bound states
in open quantum systems}
  \author{Henning Schomerus}
  \affiliation{Max-Planck-Institut f{\"u}r Physik komplexer Systeme,
  N{\"o}thnitzer Stra{\ss}e 38, 01187 Dresden, Germany}
  \author{Jakub Tworzyd{\l}o}
 \affiliation{Instituut-Lorentz, Universiteit Leiden, P.O. Box 9506,
 2300 RA Leiden, The Netherlands}
\affiliation{Institute of Theoretical Physics, Warsaw University, Ho\.{z}a 69, 00--681 Warsaw, Poland}
\date{\today}

\begin{abstract}
In the semiclassical limit of open ballistic quantum systems, we demonstrate the
emergence of instantaneous decay modes guided by classical
escape faster than the Ehrenfest time.
The decay time of the associated quasi-bound states is smaller
than the classical time of flight.  
The remaining
long-lived quasi-bound states
obey random-matrix
statistics,
renormalized in compliance with the recently proposed fractal Weyl law for open systems
[W. T. Lu, S. Sridhar, and M. Zworski, Phys.\ Rev.\ Lett.\ \textbf{91}, 154101
(2003)]. We validate our theory numerically for a model system,
the open kicked rotator.
\end{abstract}
\maketitle

Open quantum systems are described by non-self-adjoint operators
and hence have different spectral properties than closed ones.
Instead of an orthogonal set of stationary bound states at real energies one encounters
mutually non-orthogonal
quasi-bound states $\psi_n$ with complex energies $E_n$, 
which decay exponentially in time
with a uniform decay rate $\gamma_n/2=-\im E_n$.
The complex energies define the poles of the scattering matrix,
for which the random-matrix theory (RMT) provides a universal benchmark
in the case of random wave dynamics \cite{fyodorov} (for applications
to the transport in electronic nanostructures  see \cite{beenakker}).
RMT has been derived for disordered systems \cite{efetov},
while its status for
ballistic dynamics
is unsettled.
In this paper we investigate ballistic
systems that remain open in the classical limit---the
decay mechanism then corresponds to ballistic
escape of point particles into the asymptotic scattering region.
(This can be realized in clean nanostructures and in driven atoms,
provided the escape does not require tunneling.)
One goal is to uncover and quantify deviations from RMT.
We focus on the case of classically chaotic systems (our
methods are easily transferred to integrable or mixed dynamics).

We show that in approaching the classical
limit of these ballistic systems, 
the quasi-bound states 
separate into two different classes. 
The classes are discriminated by the Ehrenfest time
$\tau_{\rm Ehr}=\frac{1}{\lambda}\ln N$, up to which the escape observes
quantum-to-classical correspondence \cite{zaslavsky}.
($N$ is the number of ballistic escape channels
and $\lambda$ is of the order of
the Lyapunov exponent.)
Phase space regions with classical escape
faster than $\tau_{\rm Ehr}$ 
support instantaneous ballistic decay
with rate $\gamma_n$ larger than the inverse time of flight
through the system.
The quasi-bound states with instantaneous decay develop a large degree of
degeneracy, which is accompanied by a drastic departure from
orthogonality---the states are almost-linearly dependent.
Non-orthogonality of states has been identified \cite{lin,zworski}
as the principal obstacle 
for the formulation of Weyl laws to rigorously count the states in open
systems.  We restore the applicability of the Weyl law
and estimate 
the relative fraction
of instantaneous decay modes 
as $1-\exp(-\tau_{\rm Ehr}/\tau_{\rm
dwell})$, where
$\tau_{\rm dwell}$ is the mean dwell time.
The overwhelming majority of
long-lived quasi-bound states associated to phase space
regions with classical escape
slower than
$\tau_{\rm Ehr}$
are amenable for an
effective RMT \cite{peterprb}, in which the
dependence of the effective matrix dimension
on Planck's constant
conforms with  the recently discovered fractal Weyl laws in
classically chaotic open systems \cite{zworski}.

The motivation for our work arises from the
up-surging interest in the
quantum-to-classical correspondence of open quantum systems,
witnessed in particular over the past two years.
That quantum-to-classical correspondence invalidates the 
assumption of wave-chaotic motion and implies non-universal corrections
RMT was pointed out some while ago \cite{aleiner}.
However, only recent experiments and numerical investigations could
access the very large systems needed
in view of the
just-logarithmic scaling of $\tau_{\rm Ehr}$
with the effective system size (here given by $N$).
In stationary transport, the Ehrenfest time manifests itself
in the systematic suppression of shot noise 
\cite{agam,oberholzer,jakub} 
and in the on-set of classical
conductance fluctuations \cite{ucf}.
It also governs the impact of decoherence
on dynamical systems coupled to heat-bath
environments
(open systems of a different kind than investigated here)
\cite{garcia-mata,carvalho}.
Spectral footprints of the Ehrenfest time have only been addressed for closed systems, in terms of
the form factor of the spectral correlation function \cite{aleiner,formfactor}
and the
proximity-induced excitation gap in a
normal conductor next to a superconductor
\cite{jacquod,vavilov}. 

For simplicity,
we formulate our theory of quasi-bound states for quantum maps $F$
with one degree of freedom, operating on states $\varphi$ in 
a Hilbert space of finite dimension $M$ \cite{generalization}.
This dimension serves as the inverse of
the effective Planck's constant, $h_{\rm eff}=1/M$.
The evolution over one unit time step including loss is given by
$\varphi(t+1)=QF\varphi(t)$, where $Q$ is a projection operator of rank
$M-N$ which describes
survival in the system. 
This projection operator introduces sub-unitarity into the time
evolution, which is the equivalent to non-self-adjointness in the energy
domain.
The mean dwell time is 
$\tau_{\rm dwell}=M/N$.

For $M\to\infty$
(i.e., $h_{\rm eff}\to 0$) and fixed $\tau_{\rm dwell}$,  the system attains its
classical limit, in which it is
described by an area-preserving map ${\cal F}$
operating on a bounded two-dimensional phase space of normalized area
$1$,
while survival is described by a projection operator ${\cal Q}$
onto the complement of the openings.
For illustration see the phase space portrait of the open classical
standard map in Fig.\ \ref{fig:1}(a)
(details of this system are given below).

Quasi-bound states are defined by the
condition of quasi-stationarity
\begin{equation}
QFQ\psi_n=\mu_n\psi_n,\quad \mu_n=\exp(-i E_n),
\label{eq:eval}
\end{equation}
of the internal part of the wave function.
Since $QFQ$ is a sub-unitary operator, the eigenvalues
have modulus $|\mu_n|\leq 1$,
(i.e., lie inside the unit circle in the complex plane), and
the quasi-energies $E_n$ have a negative imaginary part.
The decay rate of a state $\psi_n$
is given by $\gamma_n = -2 \im E_n$.

\begin{figure}
\includegraphics[height=0.15\textwidth]{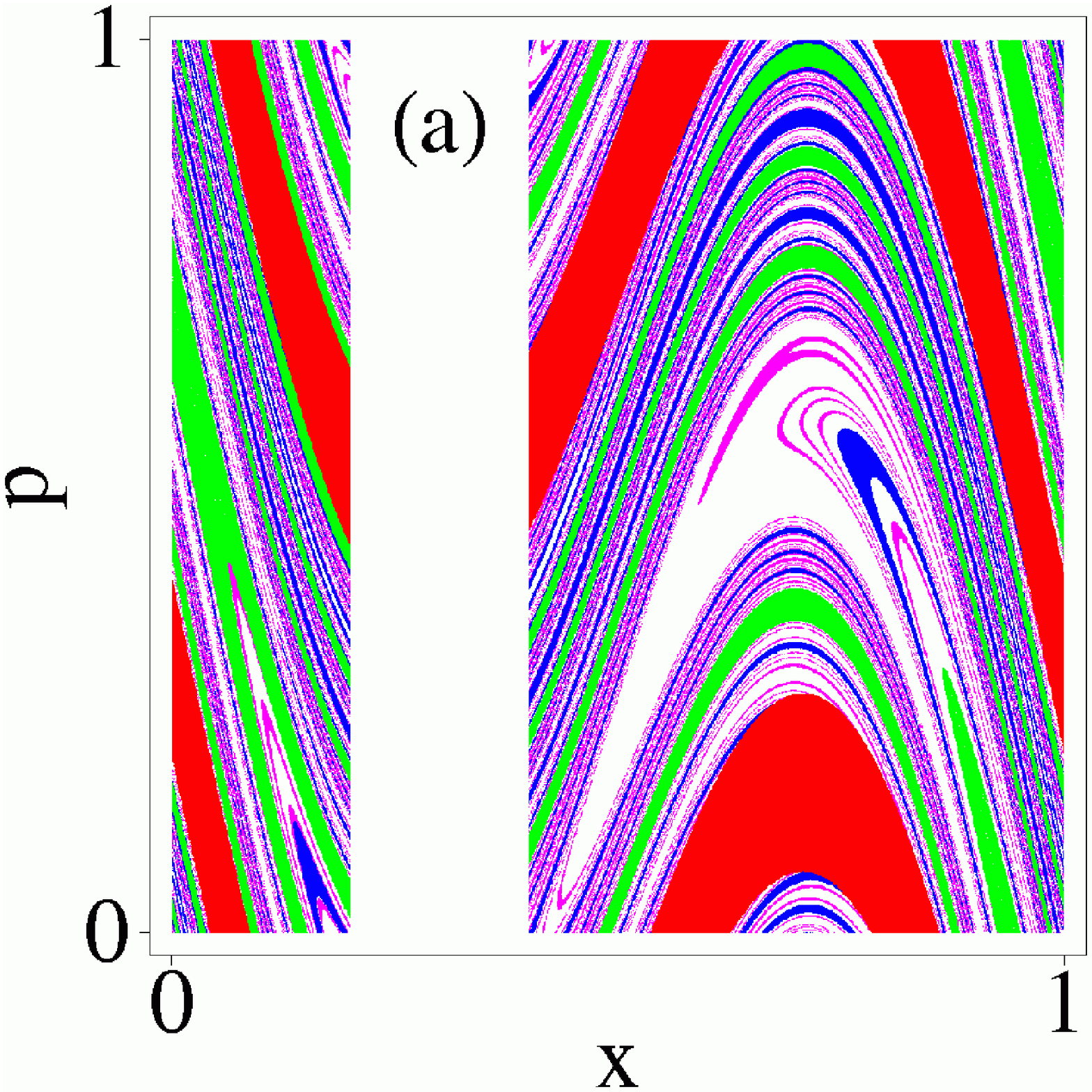}
\includegraphics[height=0.15\textwidth]{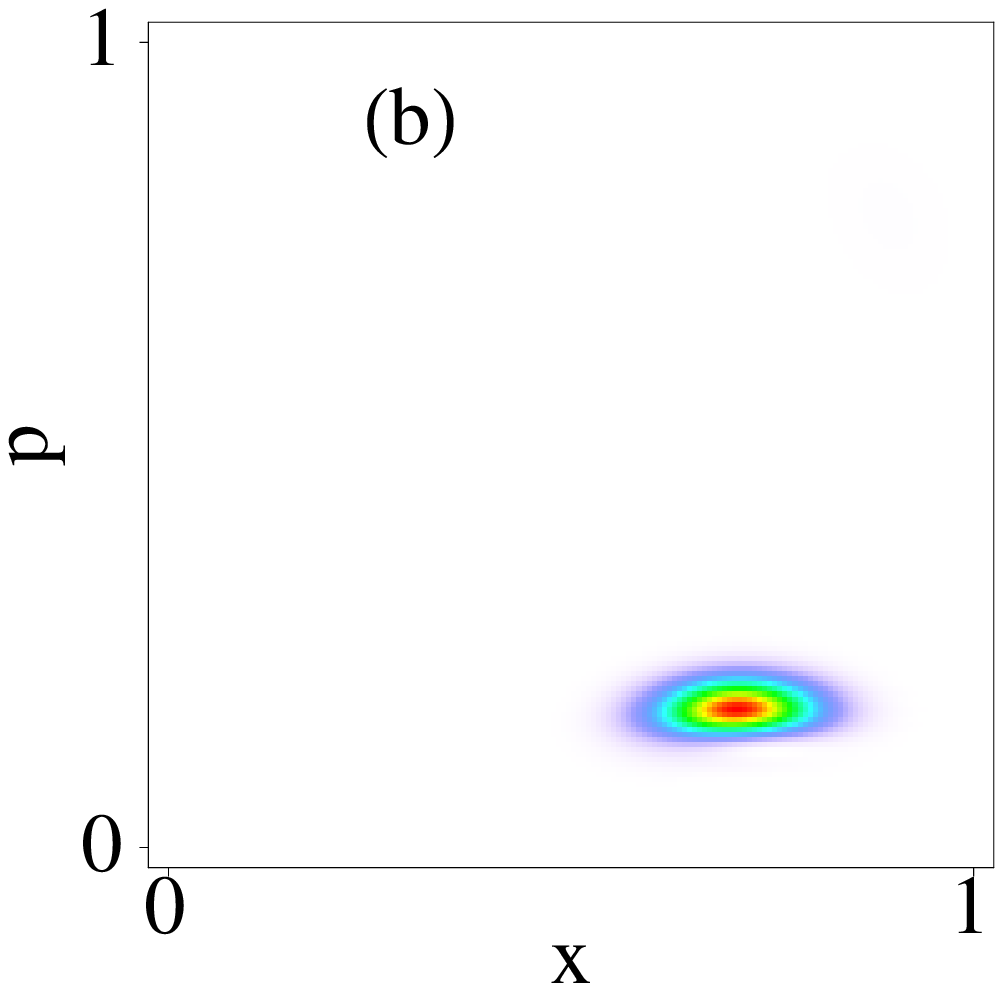}
\includegraphics[height=0.15\textwidth]{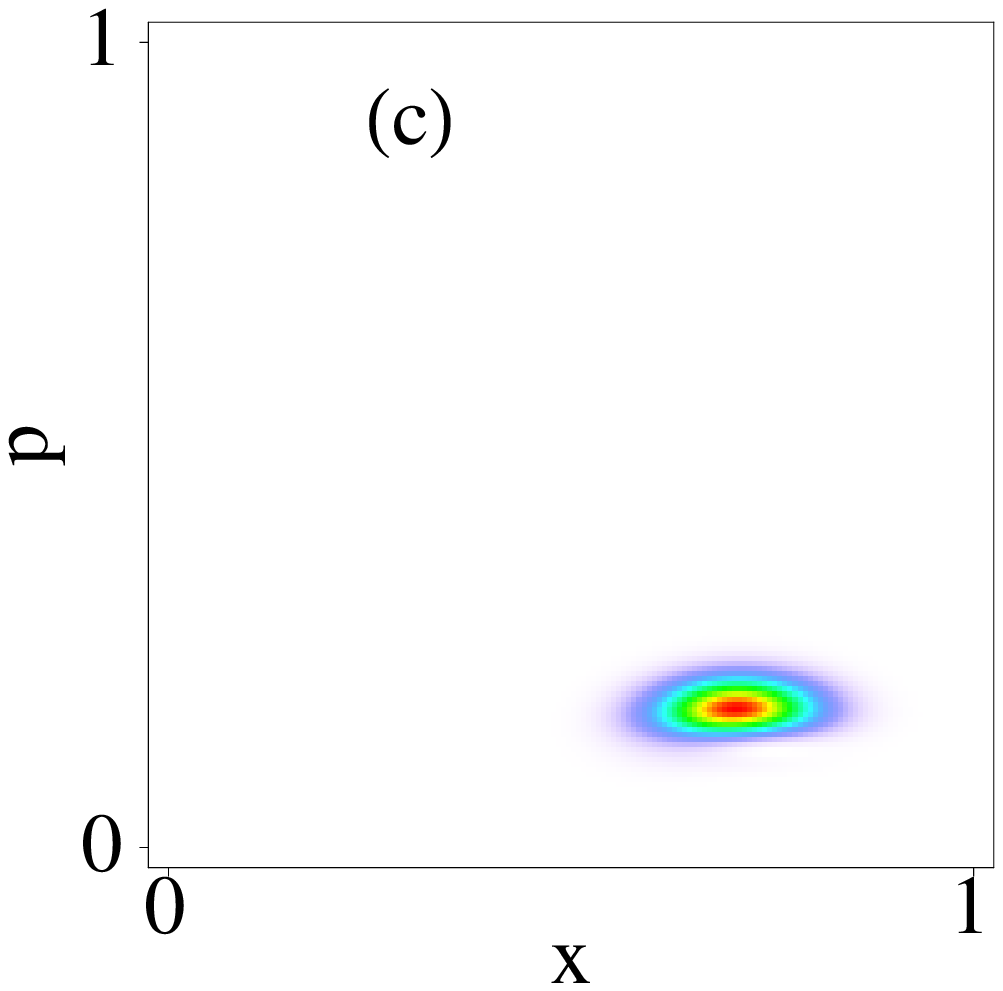}\\
\includegraphics[height=0.15\textwidth]{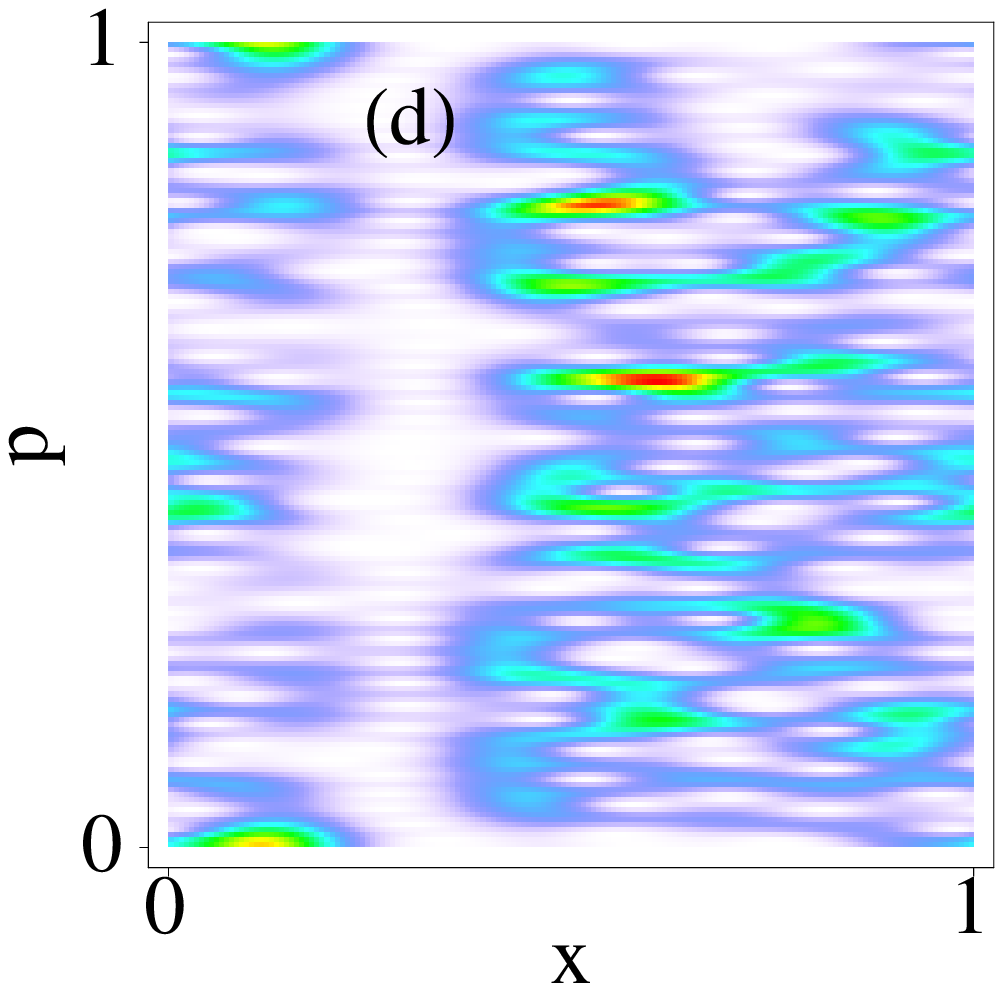}
\includegraphics[height=0.15\textwidth]{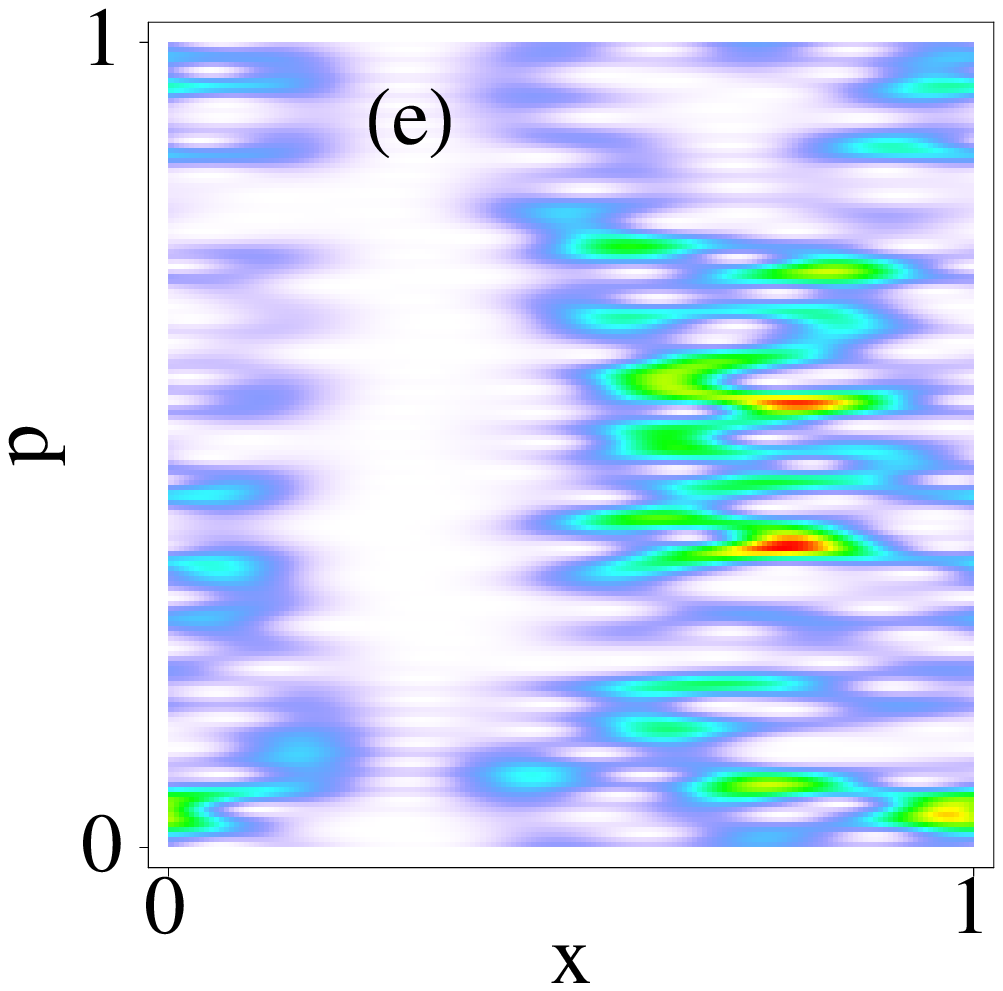}
\includegraphics[height=0.15\textwidth]{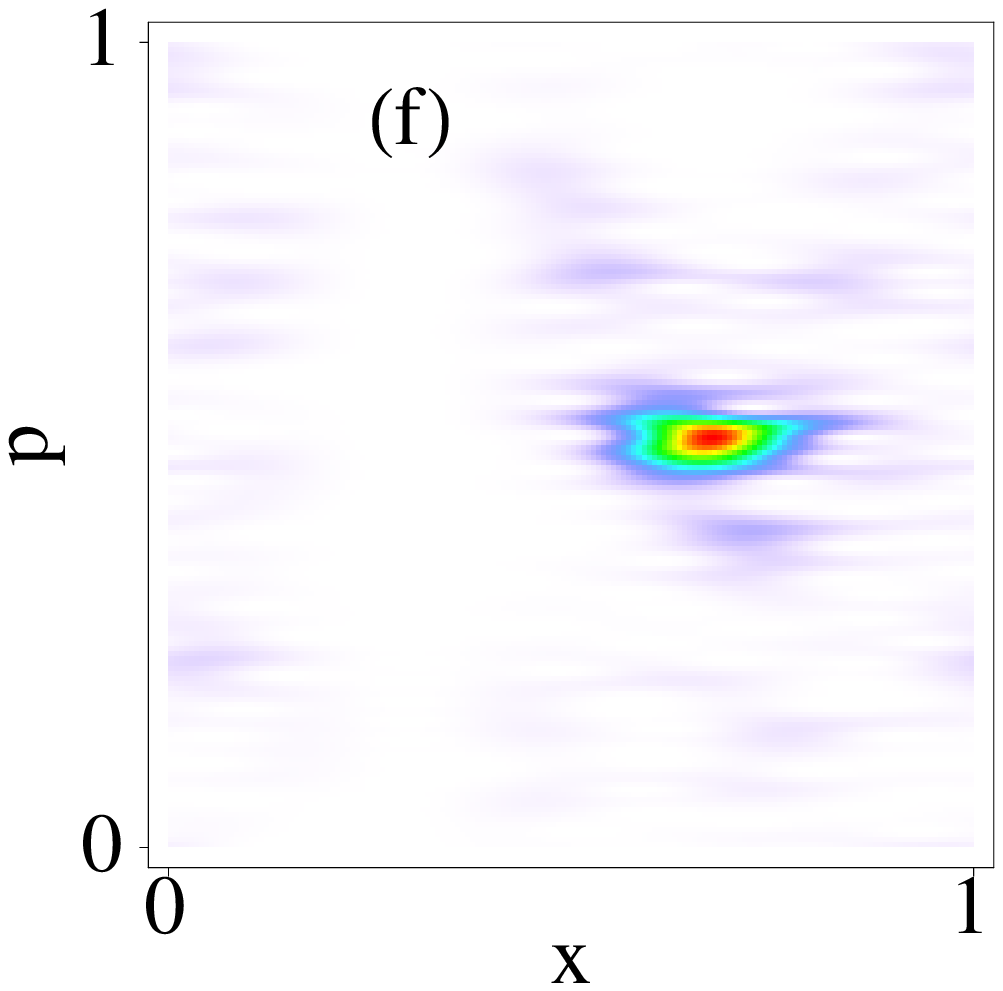}
\caption{(Color online)
(a) Regions of escape after one (red), two (green),
three (blue), four (magenta) iterations  in the open standard map (\ref{eq:smap})
with $K=7.5$ and
escape for $x\in (0.2,0.4)$ ($\tau_{\rm dwell}=5$).
The other panels show Husimi representations
of quasi-bound states for
Hilbert space dimension $M=160$.
(b,c)
Short-lived states with instantaneous ballistic decay
($|\mu|< 10^{-8}$), localized on the classical pre-image of the opening.
(d,e) Long-lived states with random-wave characteristics.
(f) Trapped long-lived state (only one state of this kind exists for
$M=160$; it has the smallest decay rate).
}
\label{fig:1}
\end{figure}

Our objective is to identify and count
instantaneous decay modes by exploring the strict
quantum-to-classical correspondence observed for escape times
shorter than the Ehrenfest time.
At the outset, note the set of $N$ trivial short-lived states in the
kernel of $Q$, all of which have eigenvalue $\mu_m=0$ ($m=1,\ldots,N$).
They can be collected into the rows of an
$M\times N$-dimensional matrix $P_0$, which fulfills
$Q=\openone-P_0P_0^T$.
The basic building blocks for the construction of a much larger number
of nontrivial short-lived
states are the
connected phase-space regions $A_{t,i}$
of escape after $t$ iterations ($t=1,2,3,\ldots$), depicted in
Fig.\ \ref{fig:1}(a).
We denote the union of all regions
with fixed $t$ by ${\cal A}_t=\cup_{i\in I_t} A_{t,i}$, where $I_t$
contains all applicable indices.
Formally associating a region  ${\cal A}_{0}$ to the opening of the
system, these regions partition phase space.

Next, we introduce the regions
${\cal B}_t=\cup_{i\in I_t'} A_{t,i}$,
where $I'_t$ restricts the index to areas larger than 
a Planck cell $h_{\rm eff}$. 
In this part of phase space
the escape of initial wave packets corresponds
to the classical particle dynamics. Consequently, we call
${\cal B}=\cup_{t<{\tau_{\rm Ehr}}}{\cal B}_t$ the region of
quantum-to-classical correspondence.
The maximal index $\tau_{\rm Ehr}$ arises because of the finite size
of the Planck cell,
and defines the Ehrenfest time of quantum ballistic escape.

In order to construct states supported by
the region ${\cal B}$ of quantum-to-classical correspondence,
let us introduce the characteristic projection
operators \cite{vallejos}
\begin{equation}
{\cal P}_t = \int_{D({\cal B}_t)}dx\,dp\,|\chi(x,p)
\rangle\langle\chi(x,p)|
,
\end{equation}
where $|\chi(x,p)\rangle$ are minimal-uncertainty wave packets 
localized at position $x$ and momentum $p$, and 
$D(\cdot)$ denotes the characteristic function of a region.
In the strict limit $h_{\rm eff}\to 0$,
the operator 
${\cal P}_t$ represents the characteristic
function of the region ${\cal A}_t$.
For finite $h_{\rm eff}$, 
only the regions ${\cal B}_t$ are well-resolved by the wave packets.
The operators
${\cal P}_t$ are defined with these smaller regions since this
guarantees
idempotency up to small corrections due to leakage.
Hence, each ${\cal P}_t$ projects onto some subspace $H_t$
of a dimension $M_t={\rm dim}\,H_t$.
This implies the representation
${\cal P}_t =  P_tP_t^T$,
where the $M$-dimensional columns of $P_t$ are mutually
orthogonal,
${\cal P}_t P_{t'}= \delta_{t\,t'}P_t$. 
We also introduce the complementary projector
$\overline{{\cal P}}=1-\sum_t{\cal P}_t=\overline{P}\,\overline{P}^T$.

Due to quantum-to-classical correspondence in the regions
${\cal B}_t$,  the semiclassical dynamics propagates states from subspace
$H_t$ to $H_{t-1}$ and finally to the opening, where they are
destroyed by $Q$:
\begin{subequations}\label{eq:qfqh}
\begin{eqnarray}
&& QFQH_t\subset H_{t-1}\quad (t>1);
\\
&& QFQH_t=\{0\} \qquad (t=0,1).
\label{eq:qfqh2}
\end{eqnarray}
\end{subequations}

Equation (\ref{eq:qfqh2}) immediately exposes the $M_0+M_1$
states in $H_0$ and
$H_1$ as quasi-bound states with eigenvalue $\mu_n=0$
(corresponding to instantaneous decay, $\gamma\to\infty$).
Does this exhaust all short-lived states?
In view of degeneracy of the associated eigenvalue
and the sub-unitarity of $QFQ$, this is by no means guaranteed.
Indeed, Eqs.\ (\ref{eq:qfqh}) naturally lead to
a partial Schur decomposition 
\cite{golub},
\begin{equation}
QFQ\approx U T U^{\dagger}
,\quad\left\{
\begin{array}{l}
U=(P_0,P_1,\ldots,P_{\tau_{\rm Ehr}},\overline{P})
,
\\
T=\left(\begin{array}{cc}
T_{11} & T_{12}
\\
0 & T_{22}
\end{array}
\right)
,
\end{array}
\right.
\end{equation}
which reveals a much larger number of small eigenvalues.
The unitary matrix $U$ leaves
the eigenvalues invariant.
The structure of the blocks of $T$
is obtained by considering the operation of $QFQ$ on the columns in $P_t$.
Equation (\ref{eq:qfqh}) implies that
$T_{11}$  is 
composed of sub-blocks that connect each subspace $H_t$ to $H_{t-1}$
(but not in the opposite direction).
Hence, this matrix is upper triangular and  all the
diagonal elements vanish. It follows that
the eigenvalue $\mu_n=0$ indeed has an algebraic 
multiplicity \cite{golub}
$M_{\rm Ehr}=\sum_{t<\tau_{\rm Ehr}} M_t$.

The difference between $M_{\rm Ehr}$ and $M_0+M_1$ can be traced back to
the break-down of the conditions for the conventional Weyl law
known for closed systems \cite{marklof},
which assumes an orthogonal set of eigenvectors and translates this into
a uniform covering of phase space.
Perturbation theory shows that
the small leakage
out of the subspaces in Eqs.\ (\ref{eq:qfqh}) lifts
the degeneracy of eigenvalues $\mu_n\approx 0$, but
the resulting $M_{\rm Ehr}$ eigenvectors are almost-linearly dependent, with
their major component confined to the subspaces $H_0$ and $H_1$.
The conventional Weyl law would underestimate their number as 
$M_0+M_1\approx M(|{\cal B}_0|+|{\cal B}_1|)$.
The Schur decomposition
carefully accounts for the admixture of the
other subspaces $H_t$ with $t\geq 2$, and provides an orthonormalized basis
$(P_0,P_1,\ldots,P_{\tau_{\rm Ehr}})$
for the short-lived states.
Equipped with such a basis, we now can formulate a new Weyl law
in analogy to the conventional
case of a closed systems: The rank $M_t$ of the
characteristic
projectors ${\cal P}_t$ can be estimated by the area covered by the regions ${\cal
B}_t$,
$ M_t\approx M|{\cal B}_t|$.
A direct consequence is
the estimate $M_{\rm Ehr}\approx M|{\cal B}|$
for the number of short-lived states.

Up to this point, 
our construction applies independent of the nature of the classical
dynamics, which may be integrable, mixed, or chaotic.
From now on we
focus on classically chaotic systems, which show a minimal amount of
system-specific details provided the dynamics are `sufficiently ergodic'
($\tau_{\rm dwell} \lambda \gg 1$).
In parallel to the considerations in Ref.\ \cite{peterprb,jakub},
a typical area $\langle |A_{t,i}|\rangle
\approx \tau_{\rm dwell}^{-1} \exp(-t \lambda)$
shrinks exponentially with time $t$
and equates to the Planck cell $h_{\rm eff}=1/M$ at
$\tau_{\rm Ehr}=\frac{1}{\lambda}\ln N$.
The number of individual regions in ${\cal B}_t$ at given $t$ is found by
balancing the exponential shrinking of areas
with the survival probability
$P(t)\sim \tau_{\rm dwell}^{-1}\exp(-t/\tau_{\rm dwell})$ for uniformly sampled
starting points in phase space.
The probability density $P(|A|)$ to initially
reside in a region of area $|A|$
then takes the power law
$P(|A|)\approx \lambda^{-1}(|A|\tau_{\rm dwell})^{-1-1/\lambda \tau_{\rm dwell}}$.
Collecting these results, the
region of quantum-to-classical correspondence
covers an area  
$|{\cal B}|\approx 1-\exp(-\tau_{\rm Ehr}/\tau_{\rm dwell})$.
We arrive at the estimate
\begin{equation}
M_{\rm Ehr}\approx M[1-\exp(-\tau_{\rm Ehr}/\tau_{\rm dwell})]
\label{eq:weyl}
\end{equation}
of the number of instantaneous decay modes in a classically
chaotic system.

The complementary region $\overline{{\cal B}}$ covers an area
\begin{equation}
|\overline{{\cal B}}|\approx \exp(-\tau_{\rm Ehr}/\tau_{\rm
dwell})=N^{-1/\lambda \tau_{\rm dwell}}.
\label{eq:b2}
\end{equation}
In this region we can expect fully developed wave chaos, described by
RMT with an
effective matrix dimension
\begin{equation}
\overline{M}=M-M_{\rm Ehr}\approx M N^{-1/\lambda
\tau_{\rm dwell}}.
\label{eq:m2}
\end{equation}
Finally,
with the  RMT prediction of
Ref.\ \cite{zyczkowski}, the number of decay modes
with $\gamma_n$ smaller than a fixed value $\gamma<1/\tau_{\rm Ehr}$ is
estimated as
\begin{equation}
n_\gamma=\overline{M}[1-\tau_{\rm dwell}^{-1}(1-e^{-\gamma})^{-1}]
.
\label{eq:fractional}
\end{equation}
For fixed $\tau_{\rm dwell}$ we obtain
$n_\gamma\propto h_{\rm eff}^d$,
hence, a power-law dependence on $h_{\rm eff}=1/M$
with non-integer exponent $d=(1/\lambda \tau_{\rm dwell})-1$.
This is precisely of the form of the fractal Weyl laws recently put forward 
by Liu et al.\ \cite{zworski},
who formulated a trace formula for
the long-lived quasi-bound states associated to the classical
repeller of a chaotic open system (while we arrived at the result
by constructing the short-lived states).

We examine our predictions for
a representative system with ballistic openings, the
open kicked rotator also used in previous investigations
\cite{jacquod,jakub,ucf} of the 
Ehrenfest time.
In the classical limit, the phase space of this system is the torus
$(x,p)\in[0,1)\otimes[0,1)$ with periodic boundary conditions.
The phase-space density
$\rho_{t+1}(x,p)={\cal QF}(\rho_t(x,p))$ evolves
in discrete time-steps
according to the standard map ${\cal F}$, specified by
\begin{eqnarray}
&& x_{t+1}=x_t+p_t+\frac{K}{4\pi}\sin2\pi x_t \quad(\mbox{mod}~1)
,
\nonumber\\
&& p_{t+1}=p_t+\frac{K}{4\pi}(\sin 2\pi x_t+\sin 2\pi x_{t+1})
\quad(\mbox{mod}~1)
,\quad~
\label{eq:smap}
\end{eqnarray}
which is
followed by leakage in terms of a projection operator ${\cal Q}$
that discards the density in a coordinate
strip $(x,p)\in {\cal A}_0=(x_0,x_0+1/\tau_{\rm dwell})\otimes [0,1)$.
For  kicking strength
$K\gtrsim 7$,
the standard map displays well-developed global chaos,
characterized by a Lyapunov exponent $\lambda\approx\ln (K/2)$.
We fix
$K=7.5$, $x_0=0.2$, $\tau_{\rm dwell}=5$.
The quantum dynamics takes place in a
Hilbert space of even integer dimension $M$, spanned by a basis of
discretized position states at $x=m/M$, $m=0,1,\ldots,M-1$.
The
time-evolution operator is
\begin{equation}
F_{mm'}=
\frac{1}{\sqrt{iM}}
e^{
\frac{i\pi}{M}
(m'-m)^2-\frac{iMK}{4\pi}[\cos \frac{2\pi m}{M} +\cos
\frac{2\pi m'}{M}]
}
,
\end{equation}
and 
$Q= {\rm diag}(\openone_{N\times N},
O_{N\times N}, \openone_{(M-2N)\times(M-2N)})$.

Husimi phase-space representations $H(x,p)=|\langle\chi(x,p)|\psi_n\rangle|^2$
of representative
quasi-bound states for $M=160$ are shown in
Fig.\ \ref{fig:1}(b-f).
The  short-lived states in panels (b,c)
have major component in the space $H_1$
associated to the region ${\cal A}_1$,
and overlap in violation of the assumptions for the conventional Weyl law.
Long-lived states (d,e) show random, delocalized wave patterns.
Exceptions are a small number of
long-lived trapped states, which are
localized in the region of slowest escape. For $M=160$ only one such
state exists, which is shown in Fig.\ \ref{fig:1}(f).

\begin{figure}
\includegraphics[width=0.4\textwidth]{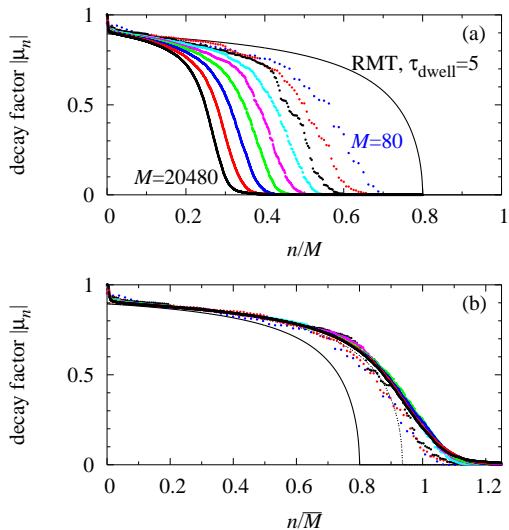}
\caption{(Color online)
Ordered decay factors $|\mu_n|=\exp(-\gamma_n/2)$ of the open kicked rotator
with
$M= 2^m\cdot 80$,
$m=0,\ldots,8$, and the other parameters as in Fig.\ \ref{fig:1}(a),
plotted as a
function of the relative index $n/M$ (panel a) or 
of the renormalized
relative index $n/\overline{M}$, where $\overline{M}$ is obtained from
Eq.\ (\ref{eq:m2})
(panel b).
The solid curve is the
RMT result \cite{zyczkowski}. The dashed curve is
this result with $\overline M$ fitted to the data.
}
\label{fig:2}
\end{figure}

Figure \ref{fig:2}(a) shows the decay factors
$|\mu_n|=\exp(-\gamma_n/2)$,
ordered by the decay rate $\gamma_n$ (starting with the
long-lived states).
The data covers a range of inverse Planck's constants between
$M=80$ and $M=20480$, and is presented as a function of the relative index
$n/M$.
In accordance with the increasing area covered by the region  ${\cal B}$
of quantum-to-classical correspondence,
the relative fraction of short-lived states increases with $M$.
Figure \ref{fig:2}(b) shows the decay factors as a function of the renormalized
relative index $n/\overline{M}$, with $\overline{M}$ estimated from 
Eq.\ (\ref{eq:m2})
and $\lambda=1.32$ approximated by the Lyapunov exponent
of the closed system.
The collapse onto one 
curve for all values of $M$
confirms the predicted scaling (\ref{eq:weyl}) of the number of
short-lived states
with the Ehrenfest time.

Also shown in Fig.\ \ref{fig:2} is the result of RMT,
 obtained by substituting the matrix $F$
with a random matrix from the circular orthogonal ensemble.
The almost perfect agreement found by fitting the effective matrix
dimension
$\overline{M}$
to the data
[dashed curve in Fig.\ \ref{fig:2}(b)] demonstrates
that effective RMT
applies to a vast majority of the long-lived
states.
Since the number of states $n_\gamma$ is obtained by inverting this 
curves, the collapse of the data onto the effective RMT curve
confirms the fractal Weyl law (\ref{eq:fractional}).

In summary, we identified instantaneous decay modes
(exceptionally short-lived quasi-bound states) in open
ballistic quantum systems, which 
capitalize on escape routes shorter than the Ehrenfest time.
The large numbers in
which these
states emerge in the semiclassical limit
is revealed only after a regularization of Weyl's law,
which is required because of the
almost-linear dependence of these states.
For chaotic classical dynamics,
the remaining long-lived states obey a RMT with an
effective matrix dimension
complying to a fractal Weyl law, from which one can conclude that
such laws
and the formation of instantaneous decay modes are intimately related.

We gratefully acknowledge enlightening
discussions with M. Zworski and S. Nonnenmacher.
J.T. acknowledges the financial support provided through the European
Community's Human Potential  Programme under contract
HPRN--CT--2000-00144,
Nanoscale Dynamics.

\end{document}